\documentclass[english,11pt]{article}

\usepackage{amssymb}
\usepackage{graphicx}
\usepackage{amsfonts}
\usepackage{bbm}
\usepackage[T1]{fontenc}
\usepackage[latin1]{inputenc}
\usepackage{babel}
\usepackage{color}

\newcommand{\be}{\begin{equation}}
\newcommand{\ee}{\end{equation}}
\newcommand{\rf}[1]{(\ref{eq:#1})}

\begin{document}

\title{ Gravitational Waves\\ from\\ Coalescing  Binary Sources}

\author{M. D. Maia\\Universidade de Brasília, 70910-900 Brasília  D.F.\\maia@unb.br\\
Essay  Written  for  the  Gravity Research  Foundation\\ 2010  Awards Essay on  Gravitation}

\maketitle

\abstract{
Coalescing  binary  systems (eg pulsars, neutron stars  and black holes) are    the most likely  sources of  gravitational radiation, yet  to be detected  on  or  near  Earth, where  the local gravitational  field is  negligible and the Poincaré   symmetry  rules.
On the other hand, the general  theory of  gravitational waves  emitted  by   axially  symmetric rotating sources predicts the existence of  a  non-vanishing  news  function. The  existence of  such  function implies  that, for  a  distant  observer,   the asymptotic   group  of   isometries, the BMS group, has  a  translational  symmetry that   depends on the  orbit   periodicity  of the source, thus  breaking the  isotropy  o the  Poincaré translations. These results  suggest the  application of  the   asymptotic   BMS-covariant  wave  equation  to obtain  a proper  theoretical  basis  for  the   gravitational  waves  observations.
 }

\newpage

\section{Gravitational Waves  from  Binary  Sources}

Coalescing  binary  systems composed by   pulsars, neutron stars  and black holes are  considered  to be the  predominant sources of  gravitational  waves to be detected  by laser interferometers   or   by  resonant mass  detectors  \cite{review1,review2}. Yet,  recent  reports from  various  research  groups  tell that  isolated or    combined operations  of these  detectors  have  failed  to  produce  any   evidences  of  gravitational waves,  although they do not  exclude  a possible  success  in the near  future \cite{results1,results2,results3,results4}.

On the theoretical  side, the most comprehensive  study  of  gravitational  waves  emitted  by axially  symmetric  rotating  systems,  which includes    binary systems,  was  presented  by   Bondi  and  collaborators  in the  early  60's.  Essentially,
the general metric   for  a  rotating  axially  symmetric  gravitational  field  is   expressed  in  spherical  coordinates $(u,\theta,\phi)$  where   $u=t-r$  is  the retarded time. In these  coordinated the  source metric  takes  the general form \cite{Bondi}
\be
ds^2 =  g_{00}du^2  +2g_{01}du dr +2g_{02}du d\theta-g_{22} d\theta^2 -g_{33}d\phi^2  \label{eq:metric}
\ee
where
\begin{eqnarray*}
&&g_{00}=   - A^2 r^2 e^{2\alpha} + \frac{B}{r}e^{2\beta},\;\;  g_{01} =e^2\beta,\;\; g_{02}=A r^2 e^{2\alpha},\;\;
g_{22}= r^2 e^2\alpha,\;\;  g_{33} =r^2sin^2 \phi \; e^{-2\alpha}
\end{eqnarray*}
and where  $A,B,\alpha,\beta$  are  functions of  $u$  and  $\theta$.  Replacing this  metric  in the vacuum  Einstein's  equations  $R_{\mu\nu}=0$  and  setting proper  boundary  conditions,  the system  can be integrated in  $r$ and  $\phi$,  with   partial  integration constants depending  on  $\theta$ and  $u$. From the  properties  of  the Riemann  geometry
and  of the  metric \rf{metric}  these partial  integration constants can  be  reduced    to  a single   function  ${\cal N}(u,\theta)$ \emph{called the news  function},   related to  the  mass  decay  of  the  gravitating system.
Taking the  expansion of the solution in powers of  $1/r$ ,    the Petrov  classification shows   that   \emph{the system  emits  gravitational  radiation if  and  only if  the  news  function does  not  vanish}:
\be
{\cal N}(u,\theta)\neq 0 \label{eq:news}
\ee
After the coalescence, when  the binary  system  collapses  into a  spherically  symmetric  configuration,  this  function vanishes \cite{Bondi,Sachs}.

Perhaps  of  greater  importance,  was  the  analysis  of the   gravitational  waves in  an  asymptotically  flat space-time,  where the  asymptotically flat   group of  isometries is appropriately  defined by the conditions
\be
R_{\mu\nu\rho\sigma}\rfloor_{r\rightarrow  \infty} =0, \;\;\;\mbox{\pounds}_{\xi}g_{\mu\nu}\rfloor_{r\rightarrow\infty}=   \xi_{(\mu ;\nu)}=0
  \label{eq:BMS}
\ee
These  equations   define the  Bondi-Metzner-Sachs (BMS) group,  a semi-direct  product of  the Lorentz  group with   the   group of  supertranslations.  The supertranslation  operators  have  the effect of breaking  the  translational  isotropy of the  Poincaré  group  in  the  sense that    \emph{they  depend on the rotation  angle  $\theta$  of  the   source}.
This  somewhat  surprising  result  produced  a
distraction to   Bondi's original  gravitational wave program,
suggesting  the possibility   that  elementary   particles  could  be  defined in a  gravitational environment, by using the  classification of the  unitary  irreducible  representations  of  the  BMS group  instead of  the  Poincar\'e group.  However, it was soon    found  that  the BMS  is  an  infinite Lie  group  whose  irreducible unitary  representations  are  impossible to classify
in any  useful  way \cite{McCarthy}.     Another pursed research  was  on  the nature of the null-cone  structure  at infinity  associated  with the BMS  symmetry \cite{NP}.

\section{The BMS-Covariant Wave Equation}

The  standard  approach  to  gravitational  waves produced by a  distant  source assumes  that the  Minkowski  metric  is perturbed by the incoming  wave  as
$g_{\mu\nu}  =  \eta_{\mu\nu}  +  \epsilon h_{\mu\nu},\;\;\; \mbox{with}\;\; \epsilon^2 <<\epsilon  $.
Replacing this in the vacuum Einstein's  equations  and  using the
 de Donder  gauge, we obtain the well  known linear  wave  equation  in  Minkowski's  space-time
\be
\Box^2_\eta  \Psi_{\mu\nu}  = \eta^{\alpha\beta}\partial_\alpha\partial_\beta \Psi_{\mu\nu}= 0,\;\; \;\; \;\;\; \Psi_{\mu\nu} =  \eta_{\mu\nu}  -\epsilon\frac{1}{2}h_{\mu\nu}  \label{eq:wave}
\ee

 On the other hand,  since the    supertranslations depend on the  source  rotation, the asymptotically flat   BMS-invariant  metric      also carry informations on that  rotation angle.

 The general expression of  the  BMS-invariant metric $\beta_{\mu\nu}$ has been  shown  to be   \emph{not invariant  under the  Poincaré  group},   except in  the particular   case  when the  source collapses into a   spherically symmetric  system \cite{Crampin}. It  follows  that instead of  \rf{wave},   the  correct   wave  equation will also be invariant  under  the  BMS  group, written   with  the BMS-invariant  metric.
In order to  find   this   equation   we  follow a  procedure  similar  to the derivation  of the  Hartle-Brill-Isaacson high frequency   gravitational  wave  equation \cite{HBI},  where the perturbation of a  generic  background  metric  $\gamma_{\mu\nu}$  by an incoming gravitational wave  is  given by
\[
g_{\mu\nu} =\gamma_{\mu\nu}   + \epsilon h_{\mu\nu}   + \cdots
\]
Defining  the    wave tensor
\[
\Psi_{\mu\nu}  =  h_{\mu\nu}  -\epsilon\frac{1}{2}h \gamma_{\mu\nu},\;\;\;\Psi=  \gamma^{\alpha\beta}\psi_{\alpha\beta}
\]
Using  again the linear   condition  $\epsilon^2  << \epsilon$
in the   vacuum  Einstein's  equations $R_{\mu\nu}(g)=0$ for the perturbed  metric,  we  obtain  the de Rham gravitational wave  equation  \cite{deRham,Zhakarov}
\[
\Box^2_\gamma \Psi_{\mu\nu}\equiv  \gamma^{\alpha\beta}\Psi_{\mu\nu;\alpha\beta}
+2\stackrel{(\gamma)}{R}_{\alpha\mu\nu}{}^\beta \Psi^{\alpha}_{\beta} +
\stackrel{(\gamma)}{R}_{\mu\alpha} \Psi^{\alpha}_{\nu}  + \stackrel{(\gamma)}{R}_{\nu\alpha} \Psi^{\alpha}_{\mu}=0
\]
describing  the  propagation of ``high frequency'' gravitational  waves  in an arbitrary  background.

In the   case of   an  asymptotically-flat  BMS-invariant metric  metric  $\beta_{\mu\nu}$  we  have
 $\stackrel{(\beta)}{R}_{\mu\nu\alpha\beta}   =0 $,
so that   the  BMS-covariant   wave  equation  becomes simply (notice  the covariant derivative  with respect  to  $\beta_{\mu\nu}$)
\be
\Box^2_\beta \Psi_{\mu\nu}\equiv
\beta^{\alpha\beta}\Psi_{\mu\nu;\alpha\beta}=0 \label{eq:deRham}
\ee
As it  is   transparent  from this  equation,  binary  systems  produce  two periodic  but  distinct   effects  on the  asymptotic detector:  The
 gravitational wave  itself  described  by the  wavefunction $\psi_{\mu\nu}$  and  the   supertranslations
included in  background metric  $\beta_{\alpha\beta}$.

Therefore, even in  an asymptotically  flat space-time
perturbed  by an incoming  gravitational wave   associated with a  binary   source,   the gravitational  wave    equation to be  considered is  the asymptotically flat   BMS-covariant  equation  \rf{deRham}.  The particular equation \rf{wave}  holds only in the ideal  case where  the  news function  are negligible, in  which case  no  gravitational  waves  are  to  be   observed.
 Of  course, the angular  dependence of  the    metric $\beta{\mu\nu}$   requires  the  knowledge of  the  most predominant binary  system affecting the detectors in the  vicinity of  the Earth. In principle this  can   be  determined  from  current  astronomical  data,  as  for  example  from  the  Sloan  Digital Sky  Survey  (SDSS),   as  well  as from  further improvements in    the  gravitational wave  detectors.

The   BMS  group  was    obtained in  an  epoch  when  the only  available gravitational  wave detector  was Weber's  original   resonant bar,  and  when the relevance of  binary  sources
 was not yet well understood.
Today,   astronomical  observations   and  the  development of  gravitational wave  detectors   have been dramatically improved,   making it  possible  to  verify the  effectiveness of  the  BMS  symmetry  to  the  gravitational  wave  astronomy.  In this  respect, gravitational  wave  laser interferometers  can be  regarded  as  a  sophisticated  adaptation of the Michelson-Morley  interferometer  to   measure gravitational  waves.  The  negative  result of that
experiment   eventually led us  to  a major   change  in our  concepts of  symmetry in  physics,   replacing the  Galilean  group  by  the Poincaré group,  which  is  consistent with  our  present understanding of  particle physics  and quantum  fields.  Likewise,
 if  binary  sources  of  gravitational waves  are predominant at the location of  an asymptotic  observer, then
the  observation of  a gravitational wave  signature
means  also  an  evidence  for  the  BMS  group,   which is effective at the very long  wavelength scale,
without  compromising the   small length  scale  of  particle  physics,   which  remain  covariant  under  the Poincaré group.

\end{document}